\documentclass[aps,prc,twocolumn,superscriptaddress,showpacs,eqsecnum]{revtex4}
\usepackage{graphicx}
\usepackage{dcolumn}
\usepackage{bm}

\newcommand{\ran}{($\alpha$,n)}

\newcommand{\rng}{(n,$\gamma$)}

\newcommand{\spro}{s-process}
\newcommand{\rpro}{r-process}

\newcommand{\lu}{$^{176}$Lu}

\newcommand{\hf}{$^{176}$Hf}
\newcommand{\loK}{low-$K$}
\newcommand{\hiK}{high-$K$}
\newcommand{\Kmix}{{\it{K}}-mixing}

\begin{document}

\title{
Properties of the $\mathbf{5^-}$ state at 839\,keV in $\mathbf{^{176}}$Lu and
the s-process branching at $\mathbf{A = 176}$
}

\author{P.\ Mohr}
\email[E-mail: ]{WidmaierMohr@t-online.de}
\affiliation{
Diakonie-Klinikum Schw\"abisch Hall, D-74523 Schw\"abisch Hall,
Germany
}

\author{S.\ Bisterzo}
\author{R.\ Gallino}
\affiliation{
Dipartimento di Fisica Generale, Universit{\`a} di Torino,
Via P.~Giuria 1, I-10125 Torino, Italy
}

\author{F.\ K\"appeler} 
\affiliation{
Forschungszentrum Karlsruhe, Institut f\"ur Kernphysik, P.O. Box 3640,
D-76021 Karlsruhe, Germany
}

\author{U.\ Kneissl}
\affiliation{Institut f\"ur Strahlenphysik, Universit\"at Stuttgart,
Allmandring 3, D-70569 Stuttgart, Germany
}

\author{N.\ Winckler}
\affiliation{GSI, Planckstra{\ss}e 1, D-64291 Darmstadt, Germany}

\date{\today}

\begin{abstract}
The \spro\ branching at mass number $A = 176$ depends on the coupling between
the \hiK\ ground state and a low-lying \loK\ isomer in \lu . This coupling is
based on electromagnetic transitions via intermediate states at higher
energies. The properties of the lowest experimentally confirmed intermediate
state at 839\,keV are reviewed, and the transition rate between \loK\ and
\hiK\ states under stellar conditions is calculated on the basis of new
experimental data for the 839\,keV state. Properties of further candidates for
intermediate states are briefly analyzed. It is found that the coupling
between the \hiK\ ground state and the \loK\ isomer in \lu\ is at least one
order of magnitude stronger than previously assumed leading to crucial
consequences for the interpretation of the \lu /\hf\ pair as an $s$-process
thermometer.
\end{abstract}


\maketitle

\section{Introduction}
\label{sec:intro}
\begin{figure}[thbp] 
  \centering
  \includegraphics[width=7.4cm]{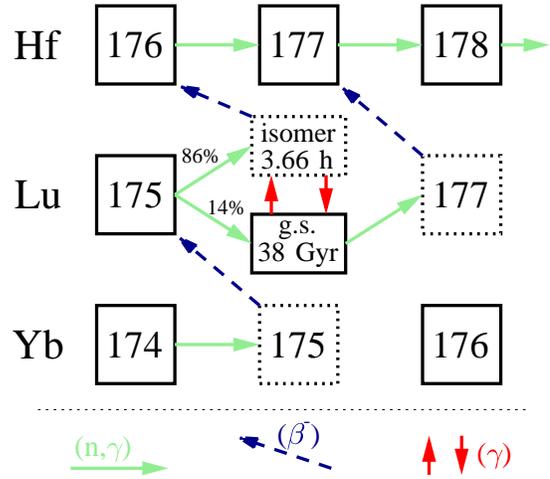}
  \caption{ 
   (Color online) Schematic view of the \spro\ branching at $A = 176$. Stable
    (unstable) nuclei are shown with full (dotted) boxes. Note that the
    long-living ground state of \lu\ is quasi-stable for the timescale of the
    \spro . Horizontal arrows indicate \rng\ neutron capture reactions, the
    dashed arrows correspond to $\beta^-$ decays, and the vertical
    arrows indicate the thermal coupling between the \hiK\ ground state and
    the \loK\ isomer in \lu . The $^{175}$Lu\rng \lu\ reaction mainly
    populates the isomer in \lu\ (86\,\%) and has only a weak 14\,\% branch to
    the ground state \cite{Heil08}.
}
  \label{fig:branch}
\end{figure}

Several studies have been devoted to the \spro\ branching at mass number $A =
176$ in the last years \cite{Heil08,Gin08,MohrPOS,Wiss06}. \lu\ and \hf\ are
$s$-only nuclei and shielded from the \rpro\ by stable $^{176}$Yb. At first
view, the abundance ratio between the unstable \lu\ and the stable \hf\ seems
to be a perfect chronometer for the \spro\ because of the long $\beta$-decay
half-life of \lu\ of about 40 giga-years. However, this ground state half-life
of \lu\ may be dramatically reduced under stellar conditions because of the
coupling of the long-living $J^\pi = 7^-; K = 7$ ground state to a low-lying,
short-living $1^-; 0$ isomer at $E_x = 123$\,keV via so-called intermediate
states (IS) at higher energies.  The $1^-$ isomer in \lu\ also $\beta$-decays
to \hf\ with a short half-life of $t_{1/2} = 3.66$\,h. The coupling depends
sensitively on temperature which turns the chronometer into a thermometer for
the helium shell flashes of AGB stars which are the commonly accepted stellar
site of \spro\ nucleosynthesis \cite{Stra06}. The \spro\ branching at \lu\ is
schematically shown in Fig.~\ref{fig:branch}.

Under \spro\ conditions
most of \lu\ ($\approx$ 86\,\%) is produced in the isomeric state in the
$^{175}$Lu\rng \lu\ reaction \cite{Heil08}. At low temperatures the coupling
between the \loK\ isomer and the \hiK\ ground state is weak, and the isomer,
i.e.\ almost all produced \lu , $\beta$-decays to \hf . At higher temperatures
the increasing coupling depopulates the isomer and produces \lu\ in its
long-living ground state, thus increasing the \lu\ abundance and the \lu \rng
$^{177}$Lu branch. At very high temperatures the production of \lu\ is reduced
again because of the repopulation of the isomer when approaching thermal
equilibrium.

Properties of candidates for IS have been carefully studied in a series of
experiments. Photon scattering \cite{Wal07} and photoactivation of
\lu\ \cite{Kn05} have been measured using bremsstrahlung at the Stuttgart
dynamitron, and earlier experiments with bremsstrahlung have been performed
using medical and technical electron accelerators
\cite{Carr89,Carr91,Lak91,Lak95a} which are known to provide high photon
intensities \cite{Mohr07NIM}. Activation after Coulomb excitation was studied
at the Tandem accelerator at IPN, Orsay \cite{Van00}, activation using
positron annihilation was measured at the Kyoto research reactor \cite{Wat81},
and photoactivation experiments with various radioactive sources were reported
in \cite{Ver70,Nor85,Lak91,Lak95b}.  High-resolution gamma-spectroscopic
studies were performed using the $^{175}$Lu\rng \lu\ reaction at the GAMS
spectrometer at ILL, Grenoble \cite{Doll99,Klay91a,Klay91b}, using HPGe
detectors \cite{Pet92,Klay91a}, and using the $^{176}$Yb(p,n)\lu\ reaction at
the Berkeley cyclotron \cite{Les91}. The $^{177}$Hf(t,$\alpha$)\lu\ reaction
was studied at Los Alamos using a Q3D magnetic spectrograph \cite{Dew81}, and
the $^{175}$Lu(d,p)\lu\ reaction was analyzed using the Q3D spectrograph at
the Munich tandem accelerator \cite{Klay91a}. Obviously, the \spro\ branching
at $A = 176$ depends also on the neutron capture cross sections. These cross
sections have been measured with high accuracy, see \cite{Heil08,Wiss06} and
references therein, and will not be analyzed here again.

This paper focuses on experimentally confirmed properties of \lu\ and the
well-established IS at an excitation energy of $E_x = 839$\,keV with $J^\pi; K
= 5^-;4$. It will be shown that the coupling between the \hiK\ $7^-;7$ ground
state and the \loK\ $1^-;0$ isomer in \lu\ via the IS at 839\,keV is
significantly stronger than adopted in the latest analysis \cite{Heil08}. A
further enhancement of the coupling has been suggested from \Kmix\ of two
almost degenerate $7^-$ states at $E_x = 725$\,keV \cite{Gin08}.

This paper is organized as follows: In Sect.~\ref{sec:rates} the formalism for
the coupling of \loK\ and \hiK\ states in a stellar photon bath will be
reviewed. This paper focuses on experimental results for the IS at 839\,keV in
Sect.~\ref{sec:data}; here conclusions can be drawn which are firmly based on
experimental data. Special attention will be given to a recent photoactivation
experiment \cite{Kn05}. Contrary to these experimentally based results, the
recent analysis of \Kmix\ for the two $7^-$ states at 725\,keV \cite{Gin08}
has to rely on theoretical considerations. The interpretation of these
experimental results for the 839\,keV state and the translation from results
under laboratory conditions to stellar conditions will be given in
Sect.~\ref{sec:stellar}. Sect.~\ref{sec:other} lists further candidates for
low-lying IS. Astrophysical consequences will be discussed in
Sect.~\ref{sec:astro}. Finally, conclusions are given in Sect.~\ref{sec:conc}.

\section{Stellar transition rates between \loK\ and \hiK\ states}
\label{sec:rates}
Direct transitions between \loK\ and \hiK\ states are highly suppressed by
$K$-selection rules. Typically, one finds suppression factors $F_W =
\tau(\rm{exp})/\tau(\rm{W.u.})$ of the order of 100 per degree $\nu$ of
$K$-suppression ($\nu = | \Delta K - \cal{L}|$) where $\tau(\rm{W.u.})$ is the
Weisskopf estimate for the lifetime $\tau$ of an E$\cal{L}$ or M$\cal{L}$
electromagnetic transition \cite{Loeb68}.

It has been shown that thermal equilibrium is achieved within the \loK\ states
and within the \hiK\ states on a short timescale of by far less than one
second \cite{Gin08,Ward80} which is much shorter than typical timescales in
\spro\ nucleosynthesis \cite{Stra06,Gal98}, whereas transition rates between
\loK\ states and \hiK\ states are much slower and depend critically on
temperature. Consequently, \lu\ has to be considered as two different species,
one with \loK\ and one with \hiK , in nucleosynthesis calculations for the
\spro . The stellar transition rate $\lambda^\ast$ for transitions between the
\hiK\ and the \loK\ species of \lu\ is given by
\begin{eqnarray}
\lambda^\ast(T) & = & 
\int c \, n_\gamma(E,T) \, \sigma(E) \, dE \nonumber \\
& \approx &
c \sum_i n_\gamma(E_{IS,i},T) \, I_\sigma(E_{IS,i})
\label{eq:lam}
\end{eqnarray}
with the thermal photon density 
\begin{equation}
n_\gamma(E,T) = 
  \left( \frac{1}{\pi} \right)^2 \,
  \left( \frac{1}{\hbar c} \right)^3 \,
  \frac{E^2}{\exp{(E/kT)} - 1}
\label{eq:planck}
\end{equation}
and the energy-integrated cross section $I_\sigma$ for an IS at excitation
energy $E_{IS}$
\begin{eqnarray}
I_\sigma & = & \int \sigma(E) \, dE 
= \frac{2J_{IS}+1}{2J_0+1} \,
\left(\frac{\pi \hbar c}{E_{IS}}\right)^2 \, \times \nonumber \\
 & & \, \, \times \, \frac{\Gamma_{IS \rightarrow
    {\rm{low-}}K}\, \Gamma_{IS \rightarrow {\rm{high-}}K}}{\Gamma}
\label{eq:isig}
\end{eqnarray}
$\Gamma_{IS \rightarrow{\rm{low-}}K}$ and $\Gamma_{IS
  \rightarrow{\rm{high-}}K}$ are the total decay widths from the IS to
\loK\ and to \hiK\ states (including all cascades), $\Gamma = \Gamma_{IS
  \rightarrow{\rm{low-}}K} + \Gamma_{IS \rightarrow{\rm{high-}}K}$ is the
total decay width, $J_{IS}$ and $J_0$ are the spins of the IS and the initial
state, and the energy $E_{IS}$ is given by the difference between the
excitation energies of the IS and the initial state: $E_{IS} = E_x(IS) - E_0 =
E_x(IS)$ for the transition rate $\lambda^\ast$ from \hiK\ to
\loK\ states. The factor $\Gamma_{IS \rightarrow{\rm{low-}}K} \times
\Gamma_{IS \rightarrow{\rm{high-}}K} / \Gamma$ in Eq.~(\ref{eq:isig}) may also
be written as $b_{IS \rightarrow{\rm{low-}}K} \times b_{IS
  \rightarrow{\rm{high-}}K} \times \Gamma$ or $b_{IS \rightarrow{\rm{low-}}K}
\times b_{IS \rightarrow{\rm{high-}}K} \times \hbar/\tau$ where $b_{IS
  \rightarrow{\rm{low-}}K}$ and $b_{IS \rightarrow{\rm{high-}}K}$ are the
total decay branchings of the IS and $\tau = \hbar/\Gamma$ is its
lifetime. 

The total stellar transition rate in Eq.~(\ref{eq:lam}) is given by the sum
over all IS; however, from the exponential dependence of the photon density in
Eq.~(\ref{eq:planck}) it is obvious that only very few low-lying IS -- and
often only the lowest IS -- dominate the stellar transition rate. It will be
shown that the experimentally confirmed IS at 839\,keV plays a key role which
may be superseded if there is strong \Kmix\ between the two $7^-$ states at
725\,keV \cite{Gin08}.

Finally, it has to be noted that the principle of detailed balance applies
under stellar conditions to the transition rates from \hiK\ to \loK\ states
and its inverse from \loK\ to \hiK\ states:
\begin{equation}
\frac{\lambda^\ast({\rm{low-}}K \rightarrow IS \rightarrow
  {\rm{high-}}K)}{\lambda^\ast({\rm{high-}}K \rightarrow IS \rightarrow
  {\rm{low-}}K)} \approx \frac{2J_{\rm{g.s.}}+1}{2J_{\rm{iso}}+1} \,
\exp{(E_{\rm{iso}}/kT)} 
\label{eq:bal}
\end{equation}
with $J_{\rm{g.s.}} = 7^-$ (\hiK ), $J_{\rm{iso}} = 1^-$ (\loK ), and
$E_{\rm{iso}} = 123$\,keV for \lu . This ratio of reaction rates in
Eq.~(\ref{eq:bal}) is independent of the properties of the IS. Under typical
stellar conditions of the \spro\ the transition rate from the \loK\ isomer at
123\,keV to the \hiK\ ground state is much larger than its inverse rate which
populates the isomer because $(2J_{\rm{g.s.}}+1)/(2J_{\rm{iso}}+1) = 5$ for
\lu\ and $\exp{(E_{\rm{iso}}/kT)} \gg 1$ at typical \spro\ temperatures. E.g.,
at $kT = 23$\,keV -- typical for the $^{22}$Ne\ran $^{25}$Mg neutron source
during helium shell flashes -- a ratio of about 1000 is found. For the lower
temperature of about 8\,keV -- typical for the $^{13}$C\ran $^{16}$O neutron
source -- this ratio is even higher. However, at this low temperature the
reaction rate is negligibly small, and the \loK\ states and the \hiK\
states in \lu\ are decoupled.

\section{Available experimental data and interpretation}
\label{sec:data}
As already mentioned in the introduction, various types of experimental data
are available for the odd-odd nucleus \lu . In particular, because of the
stable neighboring $^{175}$Lu it is possible to use the \rng\ reaction for a
detailed spectroscopic study which has been performed at the ILL using the
ultra-high resolution spectrometer GAMS \cite{Klay91a,Klay91b,Doll99}. In
addition, neutron transfer in the $^{175}$Lu(d,p)\lu\ reaction has been used
in combination with a high-resolution Q3D magnetic spectrograph
\cite{Klay91a}, and the $^{176}$Yb(p,n)\lu\ reaction has been combined with
high-resolution spectroscopy of HPGe detectors \cite{Les91}. The obtained
information is compiled in the ENSDF online data base \cite{ENSDF} which is
based on \cite{NDS}. Further information on IS can be deduced from a Coulomb
excitation and activation experiment \cite{Van00} which has detected the
activity of the isomer in \lu\ recoil nuclei after Coulomb excitation by a
$^{32}$S projectile, and from various photoactivation experiments using
bremsstrahlung and radioactive sources
\cite{Ver70,Carr89,Carr91,Lak91,Lak95a,Nor85,Lak95b}.  Finally, a
photoactivation experiment has been performed \cite{Kn05} using the high
photon flux of the bremsstrahlung setup of the dynamitron accelerator at
Stuttgart \cite{Bel01}. Although a final analysis of these data is
unfortunately not available, the data analysis in \cite{Kn05} is sufficient to
derive the integrated cross section of the IS at 839\,keV unambiguously in
combination with the other available data \cite{Klay91a,Klay91b,Doll99,Van00}.

\begin{figure}[thbp] 
  \centering
  \includegraphics[width=7.4cm]{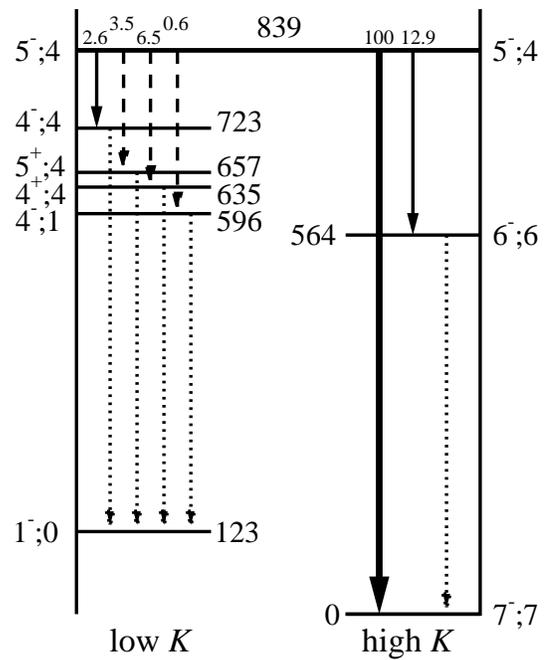}
  \caption{ 
   Partial level scheme of \lu\ with \loK\ states on the left and \hiK\ states
   on the right. The IS at 839\,keV decays to \loK\ and to
   \hiK\ states. Observed $\gamma$-ray lines are indicated by solid
   arrows. Dashed lines show tentative decays of the IS at 839\,keV. Relative
   $\gamma$-ray branches $b^\gamma_{\rm{rel}}$ normalized to the dominating
   ground state branching $b^\gamma_{\rm{rel}}(839 \rightarrow 0) = 100$ are
   given for the IS at 839\,keV. The dotted lines represent cascade
   transitions to the \hiK\ $7^-;7$ ground state and to the $1^-;0$ isomer at
   123\,keV which is the lowest \loK\ state.
}
  \label{fig:level}
\end{figure}

The integrated cross section for the transition between \loK\ and \hiK\ states
depends on the lifetime $\tau$ and the branchings $b$ of the IS, see
Eq.~(\ref{eq:isig}) in Sect.~\ref{sec:rates} and the following text. Thus, two
different approaches have been followed to determine the integrated cross
section $I_\sigma$ of the IS at 839\,keV. For illustration, a partial level
scheme of \lu\ is shown in Fig.~\ref{fig:level}.

In a first approach, see Sect.~\ref{sec:exp1},
high- and ultra-high $\gamma$-spectroscopic studies have
been performed to measure $\gamma$-ray branching ratios. This first approach
clearly identifies IS from the measured branching ratios $b$ to \loK\ and
\hiK\ states; however, an additional measurement of the lifetime $\tau$ of the
IS has to be performed which is difficult for the relevant lifetimes of the
order of several picoseconds. Lifetime measurements are further complicated by
feeding in the complex decay scheme of the heavy odd-odd nucleus
\lu\ \cite{Pet92}. It turns out that this first approach is ideal for the
identification of IS, but of limited applicability for the determination of
integrated cross sections.

In a second approach, see Sect.~\ref{sec:act},
photoactivation experiments have been performed. The
photoactivation yield is directly proportional to the integrated cross section
$I_\sigma^{\rm{lab}}$ of the contributing IS, i.e.\ this measurement provides
integrated cross sections $I_\sigma$ but does not provide the ingredients for
$I_\sigma$ which are the branching $b$ and the lifetime $\tau$. Because of the
typically broad spectral shape of the incoming photons it is difficult to
assign the photoactivation yield directly to a particular IS. This argument
holds for experiments with radioactive sources, bremsstrahlung, and virtual
photons in Coulomb excitation.

It will be shown in Sect.~\ref{sec:comb} that the combination of the
high-resolution $\gamma$-spectroscopic studies (first approach) and
photoactivation results (second approach) allows the first unique
determination of the integrated cross section $I_\sigma$ of the IS at
839\,keV. The combined analysis leads to a reduction of the uncertainty of
$I_\sigma$ from about two orders of magnitude to about a factor of two.

\subsection{Photoactivation at the Stuttgart dynamitron}
\label{sec:stutt}
The recent photoactivation experiment performed at the Stuttgart dynamitron
accelerator plays an essential role in this analysis. We briefly review the
results given only in conference proceedings up to now \cite{Kn05}. Details of
the experiment will be published elsewhere \cite{Kn09}.

\begin{figure}[thbp] 
  \centering
  \includegraphics[bbllx=70,bblly=125,bburx=560,bbury=440,width=8.0cm]{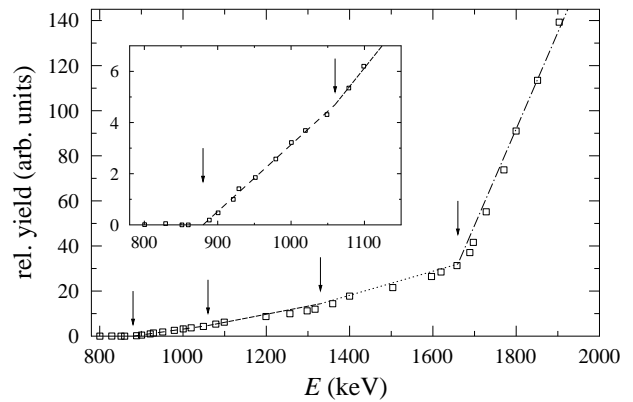}
  \caption{
   Experimental yield of the photoactivation of \lu\ as a function of the
   endpoint energy $E$. Kinks in the yield curve correspond to IS and are
   marked by vertical arrows. Between two IS the yield increases almost
   linearly, and the slope of the lines is proportional to the integrated
   cross section $I_\sigma$. The inset shows the low-energy region with the
   lowest IS at $880 \pm 30$\,keV and the second IS at $1060 \pm 30$\,keV. The
   lines between IS can be used to estimate the relative values of the
   integrated cross sections $I_\sigma$.
}
  \label{fig:yield}
\end{figure}

The high-current dynamitron accelerator at the IfS Stuttgart provides high
bremsstrahlung intensities up to energies of 4\,MeV and allows for very
sensitive photoactivation studies \cite{Bel99,Bel02}. Details of the
accelerator and the experimental set-up are given in \cite{Bel01}. Lutetium
samples with natural isotopic composition and masses of about 9\,mg of
\lu\ (total mass about 350\,mg) were irradiated with bremsstrahlung at
endpoint energies from about 800\,keV to 3\,MeV. Small steps in the endpoint
energy were used to assign the photoactivation yield to particular IS which is
otherwise a serious limitation of photoactivation experiments (see above,
``second approach''). The isomer population via IS was detected by decay
$\gamma$-spectroscopy of the 88\,keV line from the \lu $^{\rm{m}} \rightarrow
^{176}$Hf decay at higher endpoint energies. However, a practically constant
background in this $\gamma$-ray line is found from the decay of the long-living
\lu\ ground state. To improve the sensitivity, $\beta$-spectroscopy has been
used to detect electrons with energies above the endpoint of the \lu\ ground
state decay. From both $\beta$- and $\gamma$-spectroscopy data the half-life
of the isomer could be determined with $t_{1/2} = 3.63 \pm 0.02$\,h
($\gamma$-spectroscopy) and $3.640 \pm 0.004$\,h ($\beta$-spectroscopy); these
results are in good agreement with the adopted half-life of $3.664 \pm
0.019$\,h \cite{ENSDF}. The measured events in the $\gamma$- and
$\beta$-detectors have been properly corrected taking into account the decay
of \lu $^{\rm{m}}$ during the irradiation, the short waiting time between
irradiation and counting, and during the counting time. The derived
experimental yield is shown in Fig.~\ref{fig:yield} in arbitrary units.

The experimental photoactivation yield from a particular IS is proportional to
the integrated cross section $I_\sigma$ and increases almost linearly with
the endpoint energy of the bremsstrahlung. As long as the endpoint energy
remains below the lowest IS, the experimental yield obviously vanishes. When
the first IS is reached, the experimental yield starts to increase roughly
linearly. Whenever the endpoint energy exceeds a further IS, the slope of the
yield curve increases because of the additional contribution of this IS. A
schematic view of these properties of the yield curve is shown in Fig.~1 of
\cite{Bel01}.

The experimental yield is shown in Fig.~\ref{fig:yield} for endpoint energies
between 800\,keV and 2\,MeV. A clear evidence for IS is found from the kinks
of the yield curve which are located at 880\,keV, 1060\,keV, 1330\,keV, and
1660\,keV. The uncertainty of the energies of the IS is about 30\,keV which is
composed mainly of the uncertainty of the experimental endpoint energies and
of the finite energy steps in the yield curve in Fig.~\ref{fig:yield}. This
limited energy resolution complicates the assignment of the kinks in the yield
curve to individual IS in \lu . The energy difference between the lowest IS
and the second IS is $\Delta E = 180 \pm 15$\,keV. The smaller uncertainty of
about 15\,keV for the energy difference can be obtained because the error in
the endpoint assignment is approximately the same for all measurements between
800\,keV and 1200\,keV. 

From the changes of the slope in the yield curve between IS (see
Fig.~\ref{fig:yield}) a rough estimate for the relative strength of the
integrated cross sections $I_\sigma$ for the lowest IS can be derived. The
cross sections behave like $1 : 0.4 : 0.7 : 14.3$ for the first four IS in
\lu\ at 880\,keV, 1060\,keV, 1330\, keV and 1660\,keV.

\subsection{High-resolution $\gamma$-spectroscopic data (``first approach'')}
\label{sec:exp1}
The dominating decay of the IS is the direct E2 transition from the $5^-$ IS
at 839\,keV to the $7^-$ ground state. Firmly assigned is the intraband
transition from the $5^-$ IS to its $4^-$ band head of the $K = 4$ band at
723\,keV. This $4^-$ state decays via a $\gamma$-ray cascade down to the
$1^-;0$ isomer at 123\,keV (see Fig.~\ref{fig:level}). 
Thus, the properties of the $5^-$ state at
839\,keV as IS are experimentally clearly confirmed. Further, there are
several tentative assignments for transitions from the $5^-$ IS at 839\,keV to
\loK\ states which finally end at the $1^-$ isomer at 123\,keV. Depending on
these tentative assignments, the $\gamma$-ray branching to the \loK\ isomer is \cite{Klay91a}
\begin{equation}
0.021 \, \le \, b^{\gamma}_{IS \rightarrow{\rm{low-}}K} \, \le \, 0.105
\label{eq:bgamma}
\end{equation}
Using these $\gamma$-ray branchings and theoretically estimated conversion
coefficients \cite{BRICC}, this leads to
\begin{equation}
0.065 \, \le \, b^{\gamma+CE}_{IS \rightarrow{\rm{low-}}K} \, \le \, 0.149
\label{eq:bgammaCE}
\end{equation}
for the branching $b^{\gamma+CE}$ including conversion electrons in neutral
\lu . The branching under stellar conditions $b^\ast$ will be in between the
lower limit of the $\gamma$-ray branching $b^\gamma$ and the upper limit of
the branching including conversion electrons $b^{\gamma+CE}$:
\begin{eqnarray}
b^{\gamma}_{\rm{min}} \, & \le b^\ast \, & \le \, b^{\gamma+CE}_{\rm{max}} \\
0.021 \, & \le b^\ast \, & \le \, 0.149
\label{eq:bast}
\end{eqnarray}
The precise value of $b^\ast$ depends on the degree of ionization at a given
stellar temperature $T$. It can be estimated using the formalism in
\cite{Str32} which leads to about $n_{\rm{K}}^\ast \approx 0.4$ electrons in the
K-shell at $kT = 23$\,keV instead of $n_{\rm{K}} = 2$ for neutral atoms.

It is obvious that the low-energy bremsstrahlung in the photoactivation
experiment \cite{Kn05} is not able to produce highly ionized \lu . The same
argument holds for all photoactivation experiments with radioactive sources
\cite{Ver70,Nor85,Lak91,Lak95b} and for the activation experiment after
Coulomb excitation where the recoiling \lu\ nucleus has relatively low
energies. Thus, activation experiments in the laboratory determine
$b^{\gamma+CE}$ instead of $b^\gamma$ or $b^\ast$ (see also
Sects.~\ref{sec:act} and \ref{sec:stellar}). Note that the conversion
coefficient for low-energy transitions is mainly defined by the strongest
bound K-shell electrons.

It has been attempted to measure the lifetime $\tau$ of the IS at 839\,keV in
\cite{Klay91a,Klay91b,Doll99} which is mainly defined by the strongest
transition from the IS to the ground state. An upper limit could be obtained
from the measured time distributions using the generalized centroid-shift
method \cite{Klay91a}, and a lower limit was obtained from the $\gamma$-ray
induced Doppler broadening (GRID) technique \cite{Doll99}. The combined result
is
\begin{equation}
10\,{\rm{ps}} \, \le \tau \, \le \, 433\,{\rm{ps}}
\label{eq:tau}
\end{equation}
Note that the half-life $t_{1/2} \le 300$\,ps has been determined in the
original work \cite{Klay91a} which is later cited -- probably in error -- as
lifetime $\tau \le 300$\,ps \cite{Heil08,Klay91b,Doll99}. A recent photon
scattering experiment \cite{Wal07} did not see the IS at 839\,keV; from the
experimental limits $\tau \ge 1.5$\,ps could be derived which does not further
restrict the allowed range for the lifetime $\tau$ in Eq.~(\ref{eq:tau}).

The combination of the branching ratios $b^\gamma$, $b^{\gamma+CE}$, and
$b^\ast$ in Eqs.~(\ref{eq:bgamma}), (\ref{eq:bgammaCE}), and (\ref{eq:bast}),
and the lifetime $\tau$ in Eq.~(\ref{eq:tau}) leads to the following
integrated cross sections
\begin{eqnarray}
12.5\,{\rm{meV\,fm}}^2 \, & \le \, I_\sigma^{\rm{f.i.}} \, & \le \,
2479\,{\rm{meV\,fm}}^2 
\label{eq:isigion} \\
37.1\,{\rm{meV\,fm}}^2 \, & \le \, I_\sigma^{\rm{lab}} \, & \le \,
3344\,{\rm{meV\,fm}}^2 
\label{eq:isiglab}
\end{eqnarray}
with the value $I_\sigma^{\rm{lab}}$ calculated for neutral \lu\ using the
branching $b^{\gamma+CE}$ including conversion electrons, and the value
$I_\sigma^{\rm{f.i.}}$ for fully ionized \lu\ using the $\gamma$-ray branching
$b^\gamma$. Again, the stellar value $I_\sigma^\ast$ will be in between the
laboratory result and the result for fully ionized \lu .

Besides the IS at 839\,keV, further IS at higher energies have been detected
in the $\gamma$-spectroscopic experiments \cite{Klay91a,Les91}. Decay branches
to the \loK\ and to the \hiK\ part of the \lu\ level scheme have been reported
for the $5^-;4$ state at 922\,keV, the $6^-;4$ state at 963\,keV, the
$5^-$ state at 1032\,keV, and the $5^-;3$ state at 1069\,keV.
Only for the 922\,keV IS an upper limit of $\tau <
290$\,ps is available \cite{Klay91a} which translates together with the
measured branching ratios to a lower limit of the integrated cross section
$I_\sigma \gtrsim 100$\,meV\,fm$^2$. No lifetime information is available for
the IS at 963\,keV, 1032\,keV, and 1069\,keV; it is impossible to derive the
integrated cross sections $I_\sigma$ for these IS from the
$\gamma$-spectroscopic data. Furthermore, the IS at 963\,keV and 1069\,keV
cannot be seen in photoactivation because of the missing ground state branch
\cite{ENSDF}. 

Summarizing the high-resolution $\gamma$-spectroscopic experiments, a clear
assignment of IS is possible but the integrated cross section $I_\sigma$
of the IS at 839\,keV remains uncertain by about two orders of magnitude, and
$I_\sigma$ for other IS cannot be determined.

\subsection{Photoactivation experiments (``second approach'')}
\label{sec:act}
Because of the dominating ground state decay of the IS at 839\,keV it is
possible to measure the integrated cross section $I_\sigma$ by photoactivation
experiments \cite{Van00,Kn05} (see also Sect.~\ref{sec:stellar}). Other
branches from the IS at 839\,keV to \hiK\ states in \lu\ contribute only by
about 10\,\%; this marginal correction to the integrated cross section
$I_\sigma^\ast$ is neglected in the following.

Photoactivation experiments have been performed with bremsstrahlung,
radioactive sources, and virtual photons in Coulomb excitation. There is
general agreement between the various types of activation studies that the
experimental yield rises steeply with energy. This indicates that there are IS
with large integrated cross sections at higher energies. However, these IS at
higher energies are not relevant under \spro\ conditions. Consequently, it is
not possible to derive properties of low-lying IS from bremsstrahlung
photoactivation experiments with endpoint energies of several MeV
\cite{Carr89,Carr91,Lak95a}. There is also general agreement that the lowest
IS is located at energies above 650\,keV because photoactivation of \lu\ could
not be observed after irradiation with $^{137}$Cs sources ($E_\gamma =
662$\,keV) \cite{Nor85,Lak91,Lak95b} and in an experiment with bremsstrahlung
with an endpoint energy of 600\,keV \cite{Lak91}. A clear photoactivation
signal has been observed using higher energy $\gamma$-ray sources $^{60}$Co
($E_\gamma = 1332$\,keV and 1173\,keV) \cite{Ver70,Nor85,Lak91,Lak95b} and
$^{24}$Na ($E_\gamma = 2754$\,keV and 1369\,keV) \cite{Lak91}. All these
findings are nicely confirmed by the measured yield curve in
Fig.~\ref{fig:yield}.

In general, the yield in photoactivation experiments with broad incoming
photon spectra determines a weighted sum of integrated cross sections
$I_\sigma$ for all IS within the photon spectrum. The contribution of a
particular IS is weighted by the relative spectral intensity of the incoming
photons. The assignment of an experimental yield to a particular IS is thus
complicated. This holds for all photoactivation experiments under
discussion. Bremsstrahlung spectra are very broad from $E_\gamma \approx 0$ up
to the endpoint of the incoming electron beam. The primary spectra from
radioactive sources are line spectra. However, these line spectra are broadened
by Compton scattering within the heavy shielding of strong sources. Virtual
photon spectra in Coulomb excitation are also broad and can be modified by the
choice of projectile, target, and incoming projectile energy.

Effective cross sections of 45\,nb \cite{Ver70,Lak91} and 38\,nb \cite{Nor85}
have been reported for photoactivation using $^{60}$Co sources. The cross
section for $^{24}$Na irradiation is about a factor of 500 higher
\cite{Lak91}. The $^{60}$Co result of \cite{Ver70} has been translated to an
integrated cross section of $I_\sigma \approx 6 - 7$\,eV\,fm$^2$ under the
assumption of an IS located at 1\,MeV \cite{Lak91,Lak95b}. 
The photoactivation experiment after Coulomb excitation \cite{Van00} reports
$b^{\gamma+CE} \times (1 - b^{\gamma+CE}) / \tau = 12^{+10}_{-6} \times
10^{9}$\,s$^{-1}$ which translates to an integrated cross section
$I_\sigma^{\rm{lab}} = 3164^{+2637}_{-1582}$\,meV\,fm$^2$ for the strength of
IS below 1\,MeV excitation energy in \lu\ with an uncertainty of a factor of
two.

\subsection{Combination of the two approaches}
\label{sec:comb}
Although the photoactivation experiment using bremsstrahlung \cite{Kn05} does
not provide integrated cross sections yet, this experiment does provide rough
information on the excitation energies of the lowest IS from the kinks in the
measured yield curve (see Fig.~\ref{fig:yield}). The lowest IS is located at
$880 \pm 30$\,keV, and the next IS is found at $1060 \pm 30$\,keV. The energy
difference between the first and second IS is much better defined by $\Delta E
= 180$\,keV with an estimated uncertainty of about 15\,keV. The comparison
with the known IS from the $\gamma$-spectroscopic studies \cite{Klay91a,Les91}
leads to the clear conclusion that only the IS at 839\,keV and 1032\,keV with
$\Delta E = 193$\,keV have been seen in the photoactivation experiment
\cite{Kn05} whereas the state at 922\,keV does not show up in the
photoactivation yield curve. From the yield curve in Fig.~\ref{fig:yield} an
upper limit for the integrated cross section of the IS at 922\,keV can be
estimated which is about a factor of five lower than the integrated cross
section for the 839\,keV state. As will be shown below, this finding is in
agreement with the lower limit for $I_\sigma \gtrsim 100$\,meV\,fm$^2$ for
this IS from the $\gamma$-spectroscopic data.

The kinks in the yield curve of Fig.~\ref{fig:yield} at higher energies
(1330\,keV and 1660\,keV) cannot be assigned to IS from $\gamma$-spectroscopy
because the highest energies in the $\gamma$-spectroscopic studies were
1130\,keV \cite{Klay91a} and 902\,keV \cite{Les91}. A very tentative
assignment can be suggested for the kink at 1330\,keV. It may correspond 
($i$) 
to the 1332\,keV level seen in photon scattering and thus coupled to the ground
state \cite{Wal07} or 
($ii$)
to a level at 1301\,keV seen in the
$^{176}$Yb(p,n)\lu\ reaction \cite{Les91} which is coupled to the
isomer. A few levels around the IS at 1660\,keV are reported in \cite{ENSDF}
at 1617\,keV, 1655\,keV, 1679\,keV, 1689\,keV, and 1693\,keV;
however, for two of these levels not even a spin/parity assignment has been
adopted in \cite{ENSDF}.

The photoactivation experiment after Coulomb excitation \cite{Van00} is not
able to resolve individual IS. The given result in \cite{Van00} of
$b^{\gamma+CE} \times (1 - b^{\gamma+CE}) / \tau = 12^{+10}_{-6} \times
10^{9}$\,s$^{-1}$ translates to an integrated cross section
$I_\sigma^{\rm{lab}} = 3164^{+2637}_{-1582}$\,meV\,fm$^2$ for the strength of
IS below 1\,MeV excitation energy in \lu\ with an uncertainty of a factor of
two. Because the $\gamma$-spectroscopic experiments \cite{Klay91a} find only
two IS below 1\,MeV excitation energy, namely at 839\,keV and 922\,keV, the
measured activation cross section must be the sum of the cross sections of
both states. The next IS is reached at a significantly higher energy of
1032\,keV which is more than 100\,keV above the 922\,keV state and 193\,keV
above the 839\,keV state.

Because the 922\,keV state does not contribute to the photoactivation yield in
the bremsstahlung experiment, this state also cannot contribute to the Coulomb
activation yield. Thus, the integrated cross section $I_\sigma^{\rm{lab}} =
3164$\,meV\,fm$^2$ from the Coulomb excitation experiment
\cite{Van00} is dominated by the contribution of the IS at 839\,keV; more than
80\,\% of the strength in \cite{Van00} can be assigned to the 839\,keV
IS. This result is at the upper end of the allowed range of
$I_\sigma^{\rm{lab}}$ in Eq.~(\ref{eq:isiglab}) which has been determined from
$\gamma$-spectroscopy experiments \cite{Klay91a,Doll99}. Combining the limits
of the $\gamma$-spectroscopy experiments and the integrated cross section from
the Coulomb excitation and photoactivation experiments leads to the conclusion
that the branching $b^{\gamma+CE}$ must be at the upper end of the allowed
range in Eq.~(\ref{eq:bgammaCE}), and the lifetime $\tau$ of the IS at
839\,keV must be close to the lower limit of \cite{Klay91a,Doll99}. A
consistent set of parameters for the IS at 839\,keV is $b^{\gamma+CE} \approx
0.1$ and $\tau \approx 12$\,ps leading to an integrated cross section
$I_\sigma^{\rm{lab}} \approx 2000$\,meV\,fm$^2$ with an upper limit of about
3500\,meV\,fm$^2$ from the branching and the lifetime from
$\gamma$-spectroscopic studies \cite{Klay91a,Klay91b,Les91,Doll99} and a lower
limit of about 1250\,meV\,fm$^2$ from the activation experiments
\cite{Van00,Kn05}. In total, an uncertainty of a factor of two for the
integrated cross section $I_\sigma^{\rm{lab}}$ seems to be a careful realistic
estimate.

This result is also in reasonable agreement with the photoactivation using
$^{60}$Co sources with a maximum energy of $E = 1332$\,keV. The integrated
cross sections of the 839\,keV, (922\,keV), 1032\,keV, and 1330\,keV IS scale
like $1 : (\le 0.2) : 0.4 : 0.7$ (see Fig.~\ref{fig:yield} and
Sect.~\ref{sec:stutt}), i.e.\ about one half of the measured cross section of
$6 - 7$\,eV\,fm$^2$ must be assigned to the lowest IS at 839\,keV. However,
this assignment to the lowest IS has a rather large uncertainty because the IS
at $1330 \pm 30$\,keV is located very close to the $^{60}$Co energy of
1332\,keV. Additionally, a tentative state in \lu\ has been seen in photon
scattering \cite{Wal07} at 1332\,keV. If there is accidental overlap between
the primary $^{60}$Co energy and the excitation energy of an IS in \lu , then
the yield in the $^{60}$Co photoactivation experiments
\cite{Ver70,Nor85,Lak91,Lak95b} may be strongly affected by this IS at
1330\,keV.

At first view, from the allowed ranges of $I_\sigma^{\rm{lab}}$ from
$\gamma$-spectroscopy in Eq.~(\ref{eq:isiglab}) and the result
$I_\sigma^{\rm{lab}} = 3164^{+2637}_{-1582}$\,meV\,fm$^2$ from the Coulomb
activation experiment \cite{Van00} a slightly higher result of
$I_\sigma^{\rm{lab}} \approx 3000$\,meV\,fm$^2$ should be derived. However,
the limit from $\gamma$-spectroscopy is a combined limit from two independent
measurements of the branching ratio and the lifetime. A result of
$I_\sigma^{\rm{lab}} \approx 3000$\,meV\,fm$^2$ would require that both
quantities are very close to their experimental limits which is statistically
not very likely. Taking into account the relatively large uncertainties of the
Coulomb activation experiment, $I_\sigma^{\rm{lab}} \approx 2000$\,meV\,fm$^2$
seems to be a more realistic estimate for the integrated cross section of the
IS at 839\,keV in \lu . A similar result is obtained from the average of the
combined lower and upper limits from $\gamma$-spectroscopy and Coulomb
activation.

\subsection{Some further considerations}
\label{sec:further}
From the above parameters of the 839\,keV state a surprisingly strong
transition is found from the $7^-;7$ ground state to the $5^-;4$ IS at
839\,keV. The lifetime $\tau = 12$\,ps corresponds to roughly 3\,W.u.\ which
is much larger than the expected strength of about 0.01\,W.u.\ for a
$K$-forbidden E2 transition with $\Delta K = 3$ and thus $\nu = 1$. Contrary
to the 839\,keV state, the much smaller ground state transition strength for
the 922\,keV state with the same quantum numbers $5^-;4$ seems to be regular.

The strength of the transition from the $7^-;7$ ground state to
the $5^-;4$ IS at 839\,keV has also been estimated theoretically from the
measured intraband branching of the 839\,keV state to its $4^-;4$ band head at
723\,keV and the calculated strength of this intraband transition
\cite{Doll99}. It is concluded in \cite{Doll99} that ``the estimates
\ldots\ point at a lifetime of the 838.6\,keV level which is certainly below
50\,ps, probably close to the lower experimental limit of 10\,ps'' although
no explanation was found for the unusually large strength of this
transition in \cite{Doll99}.

There is one further experimental hint that the 839\,keV state couples
stronger to the \hiK\ ground state of \lu\ than the 922\,keV state. Thermal
$s$-wave neutron capture on $^{175}$Lu with $J^\pi = 7/2^+$ populates mainly
states with $J^\pi = 3^+, 4^+$ in \lu . These states which obviously cannot be
members of \hiK\ bands will preferentially decay to \loK\ states. It is found
that the number of observed $\gamma$-rays per neutron capture for the $839
\rightarrow 0$ transition is about a factor of 5.5 larger than for the $922
\rightarrow 0$ transition \cite{Klay91a}. This experimental finding further
strengthens the result that there is an unusually strong coupling of the
839\,keV IS to the \hiK\ ground state.

\section{Laboratory and stellar reaction rates}
\label{sec:stellar}
In general, dramatic changes may be found for photon-induced reaction rates
under laboratory and under stellar conditions. Whereas under laboratory
conditions the target nucleus is in its ground state (except the case of
$^{180}$Ta which is found in nature in its long-living isomeric state), under
stellar conditions excited states are populated in thermal equilibrium
according to the Boltzmann statistics. Consequently, the transition rate under
laboratory conditions depends on the direct decay width $\Gamma_{IS
  \rightarrow {\rm{g.s.}}}$ to the ground state, whereas under stellar
conditions the rate depends on the total $\gamma$-decay width to all states
which finally cascade down to the ground state; in the case of \lu\ this width
is given by $\Gamma_{IS \rightarrow {\rm{high-}}K}$ to all \hiK\ states. A
detailed discussion of the importance of thermally excited states under
stellar conditions is given in \cite{Mohr06,Mohr07}.

For the case of \lu\ the influence of thermally excited states is unexpectedly
low. The dominating IS at 839\,keV decays preferentially to the ground state,
and the transition rate under stellar conditions is only moderately enhanced
because $\Gamma_{IS \rightarrow {\rm{g.s.}}} \approx \Gamma_{IS \rightarrow
  {\rm{high-}}K}$. Compared to other uncertainties, this moderate enhancement
of the order of $10 - 20$\,\% can be neglected in this work.

However, there is another effect that has to be analyzed. The only firmly
assigned transition from the $5^-$ IS at 839\,keV to \loK\ states in \lu\ is
the intraband transition to the band head of the $K = 4$ band at 723\,keV with
a relatively small transition energy of 116\,keV. This M1 or E2 transition is
enhanced by internal conversion for neutral \lu\ atoms with conversion
coefficients $\alpha_{M1} = 2.42$ and $\alpha_{E2} = 1.93$ \cite{BRICC}. This
transition has been tentatively assigned to (M1) in \cite{Klay91a}. Because of
the relatively small difference of the M1 and E2 conversion coefficients, only
the M1 assignment is used in the following discussion.

In laboratory experiments the charge state of \lu\ depends on the experimental
conditions. Under stellar conditions \lu\ is highly ionized depending mainly
on the temperature $T$ and weakly on the electron density $n_e$
\cite{Str32}. The M1 conversion coefficient is mainly defined by the K-shell
contribution. In all experiments under study in this paper
\cite{Klay91a,Klay91b,Doll99,Les91,Pet92,Van00,Kn05} \lu\ will not be fully
ionized. This is obvious for the neutron capture experiments with thermal
neutrons \cite{Klay91a,Klay91b,Doll99} and the photoactivation with low-energy
bremsstrahlung \cite{Kn05}. But also for the
$^{175}$Lu(d,p)\lu\ \cite{Klay91a}, $^{176}$Yb(p,n)\lu\ \cite{Les91}, and
Coulomb excitation and activation experiments \cite{Van00} the energies are
not sufficient to produce fully ionized \lu . Thus, the derived transition
strengths and integrated cross sections $I_\sigma^{\rm{lab}}$, see
Eq.~(\ref{eq:isiglab}), from the activation experiments can be compared to the
$\gamma$-spectroscopic results including conversion electrons, and a combined
result $I_\sigma^{\rm{lab}} \approx 2000$\,meV\,fm$^2$ has already been given
in Sect.~\ref{sec:exp1}.

At typical stellar \spro\ conditions (thermal energy $kT = 23$\,keV, electron
density $n_e = (3-5) \times 10^{26}$\,cm$^{-3}$) the K-shell of \lu\ is
partially ionized leading to $n^\ast_{\rm{K}} \approx 0.4$ instead of
$n_{\rm{K}} = 2$ for neutral atoms. Thus, the decay width for the
\hiK\ $\rightarrow$ \loK\ transition with $E = 116$\,keV has to be reduced by
the factor $(1 + \alpha_{\rm{eff}})\,/\,(1 + \alpha_{\rm{lab}}) \approx 0.43$
with $\alpha_{\rm{eff}} \approx \alpha_{\rm{lab}} \times n^\ast_{\rm{K}}/2$
leading to an integrated cross section of $I_{\sigma}^{\ast} \approx
850$\,meV\,fm$^2$ under stellar \spro\ conditions.

The reduction factor $(1 + \alpha_{\rm{eff}})\,/\,(1 + \alpha_{\rm{lab}})
\approx 0.43$ is only valid for the low-energy transition with $E =
116$\,keV. There are further tentative assignments for transitions from the
839\,keV IS to \loK\ states with higher transition energies of $E = 181$\,keV,
203\,keV, and 243\,keV. The conversion coefficients for these transitions are
significantly smaller with $\alpha_{\rm{lab}} < 1$ compared to the 116\,keV
transition with $\alpha_{\rm{lab}} = 2.42$. The reduction factor $(1 +
\alpha_{\rm{eff}})\,/\,(1 + \alpha_{\rm{lab}})$ remains close to unity for
these tentatively assigned transitions. As the found branching $b^{\gamma+CE}$
to \loK\ states (see at the end of Sect.~\ref{sec:exp1}) is about 0.1, it is
very likely that not only the firmly assigned transition, but also some
tentatively assigned transitions contribute to the total transition strength
from the IS at 839\,keV to \loK\ states. Consequently, the reduction factor
from $I_\sigma^{\rm{lab}}$ for neutral \lu\ to the stellar $I_\sigma^\ast$ for
partially ionized \lu\ is between the minimum value of $\approx 0.43$ for the
116\,keV transition and the maximum value of unity.

Combining all the above information and its uncertainties, the final result
for the integrated cross section from \hiK\ to \loK\ states via the IS at
839\,keV under stellar \spro\ conditions is
\begin{equation}
600\,{\rm{meV\,fm}}^2 \, \le \, I_\sigma^\ast \, \le \, 2500\,{\rm{meV\,fm}}^2 
\label{eq:isigast}
\end{equation}
which is entirely based on experimental results and reliably calculated
internal conversion coefficients and stellar ionization. In short, this result
can also be given as $I_\sigma^\ast \approx 1200$\,meV\,fm$^2$ with an
uncertainty of about a factor of two. The resulting reaction rate via the
839\,keV state will be compared to other candidates for IS in the next
Sect.~\ref{sec:other}.

\section{Further candidates for low-lying intermediate states}
\label{sec:other}
\begin{figure}[thbp] 
  \centering
  \includegraphics[bbllx=15,bblly=10,bburx=375,bbury=390,width=7.4cm]{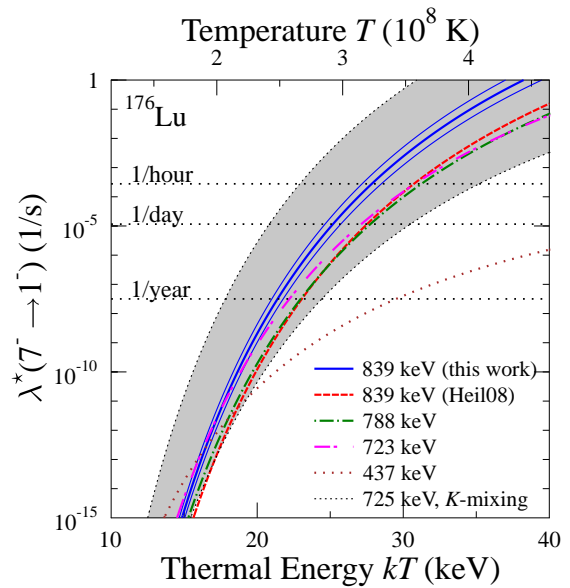}
  \caption{ 
    (Color online) Reaction rates $\lambda^\ast$ under stellar
    conditions for the transition from the \hiK\ ground state to the
    \loK\ isomer in \lu\ from Eqs.~(\ref{eq:lam}) and (\ref{eq:isig}). The
    dominating IS is located at 839\,keV; this rate is shown as thick full
    line (blue) with its uncertainties (thin full lines). A huge uncertainty
    remains for the theoretically estimated contribution from \Kmix\ of two
    almost degenerate $7^-$ states at 725\,keV \cite{Gin08}
    (thin dotted lines and gray
    shaded uncertainty). Contributions of other candidates for IS are small at
    $kT \approx 23$\,keV, i.e.\ at \spro\ temperatures. A recent
    astrophysically derived transition rate \cite{Heil08} is also shown (red
    dashed line) which is at least one order of magnitude smaller than the
    result of this work. Further discussion see text.  
}
  \label{fig:rates}
\end{figure}

The reaction rates from the \hiK\ ground state to the \loK\ isomer in \lu\ are
shown for various IS in Fig.~\ref{fig:rates}. Because of the dominance of the
IS at 839\,keV state, further candidates for IS are only briefly discussed.

In general, properties of astrophysically relevant IS should be an
intermediate $K$ quantum number and a low excitation energy. The transition
rate $\lambda^\ast$ depends linearly on the integrated cross section
$I_\sigma^\ast$ and exponentially on the excitation energy $E_{IS}$. The
discussion in this Sect.~\ref{sec:other} has to rely on theoretical
considerations because the relevant properties of most candidates for IS have
not been determined experimentally: whereas $\gamma$-spectroscopic data
\cite{Klay91a,Klay91b} clearly identify candidates for IS by their decay
branches to \loK\ and \hiK\ states, the integrated cross section $I_\sigma$
cannot be derived from experiment because of missing lifetime data. Because of
the relatively huge uncertainties of the lifetimes from theoretical Weisskopf
estimates, a further correction of the data because of weakened internal
conversion under stellar conditions -- typically of the order of a factor of
two or less -- is neglected here.

Good candidates for further IS are the band heads of the $K=4$ bands at
723\,keV and 788\,keV. Both $4^-$ states decay preferentially to a $3^-$ state
at 658\,keV which finally cascades down to the $1^-$ isomer. Both $4^-$ states
may branch to a $6^-;6$ \hiK\ state at 564\,keV although this branching has
not been observed experimentally. Assuming one Weisskopf unit for this allowed
$\Delta K = 2$ E2 transition, the integrated cross sections are $I_\sigma^\ast
\approx 10$\,meV\,fm$^2$ for the 788\,keV state and $I_\sigma^\ast \approx
2$\,meV\,fm$^2$ for the 723\,keV state. The reaction rate of both $4^-$ states
(dash-dotted lines in Fig.~\ref{fig:rates}) is significantly lower at $kT =
23$\,keV than the 839\,keV rate which is shown as thick solid line with its
uncertainties (thin solid lines). The strength of the $K$-allowed direct M3 or
E4 transition from the $4^-$ states to the $7^-$ ground state is negligible
compared to the above mentioned E2 strength.

The $5^-;0$ \loK\ state at 437\,keV decays by $\gamma$-cascades to the $1^-$
isomer. The strength of the unobserved $K$-forbidden E2 transition to the
\hiK\ ground state can be estimated from the $K$-forbiddenness $\nu = 5$
leading to $10^{-10}$\,W.u.\ for this transition and $I_\sigma^\ast = 1.1
\times 10^{-7}$\,meV\,fm$^2$. Although the 437\,keV state has a much lower
excitation energy than the other IS, this strength is by far not sufficient
for a noticable contribution to the transition rate $\lambda^\ast$ (dotted
line in Fig.~\ref{fig:rates}).

\Kmix\ of two almost degenerate $7^-$ states at 725\,keV has been studied in
detail in \cite{Gin08}. Based on the results in Tab.~I of \cite{Gin08} the
transition rate can be calculated; the thin dotted lines and the gray shaded
area show the allowed range of \cite{Gin08}. At the relevant energy of $kT =
23$\,keV the contribution of \Kmix\ may be larger or much lower than the
experimentally confirmed transition rate of the 839\,keV state. It is obvious
that the decay branches of these two $7^-$ states have to be studied
experimentally before a clear conclusion on the relevance of \Kmix\ can be
drawn. From the experimental limits of unobserved $\gamma$-transitions from
the two $7^-$ states in \cite{Klay91a} a rough estimate for an upper limit of
the integrated cross section and transition rate may be estimated which is
about one order of magnitude lower than the result for full \Kmix\ shown in
Fig.~\ref{fig:rates} \cite{Gin08,Cham08}.

Additionally, Fig.~\ref{fig:rates} shows the result of an astrophysical
determination -- see Sect.~\ref{sec:astro} -- of the transition rate of the IS
at 839\,keV with an integrated cross section $I_\sigma^\ast = 71$\,meV\,fm$^2$
\cite{Heil08} (dashed line) which is at least one order of magnitude smaller
than the experimental result derived in Sects.~\ref{sec:data} and
\ref{sec:stellar}.

Higher-lying IS have a negligible contribution to the total reaction
rate. From the combination of the $\gamma$-spectroscopic data \cite{Klay91a}
and the photoactivation yield in Fig.~\ref{fig:yield} the integrated cross
section of the IS at 922\,keV is $100\,{\rm{meV\,fm}}^2 \lesssim I_\sigma
\lesssim 500\,{\rm{meV\,fm}}^2$. Compared to the IS at 839\,keV, the 
contribution of the IS at 922\,keV is suppressed by its smaller $I_\sigma$ and
the higher excitation energy. A similar consideration for the IS at 1032\,keV
leads to an integrated cross section of about $I_\sigma \approx
1.4$\,eV\,fm$^2$ with an uncertainty of at least a factor of two. Together
with the measured branchings of the IS at 1032\,keV a lifetime of $\tau \approx
25$\,ps can be derived for this state.

\section{Astrophysical consequences}
\label{sec:astro}
\lu\ and \hf\ are produced in the so-called main component of the \spro\ which
is assigned to thermally pulsing low-mass AGB stars
\cite{Gal98,Buss99,Stra06}. The \spro\ branching at $A = 176$ is of special
interest because this branching depends very sensitively on the temperature
during \spro\ nucleosynthesis whereas most other \spro\ branchings depend on
the neutron density. 

Two neutron sources operate in thermally pulsing AGB stars. The $^{13}$C\ran
$^{16}$O reaction operates during the interpulse phase at relatively low
temperatures of about $kT \approx 8$\,keV for about $10^4$ years; it releases
about 90\,\% of the total neutron exposure. The $^{22}$Ne\ran $^{25}$Mg
reaction operates during the convective helium shell flashes at temperatures
of about $kT \approx 23$\,keV which last for about six years; the temperature
increases with the number of the thermal pulse and reaches a maximum of about
$kT \approx 27$\,keV \cite{Stra03}.

The lower temperature of about 8\,keV in the interpulse phase is not
sufficient for a thermal coupling between the \hiK\ ground state and the
\loK\ isomer in \lu . The reaction rate $\lambda^\ast$ drops below about
$10^{-15}$/s $\approx 3 \times 10^{-8}$/y already above 12\,keV (see
Fig.~\ref{fig:rates}) and is thus neglibible at 8\,keV. The \hiK\ ground state
and the \loK\ isomer have to be treated as two fully separated species in this
interpulse phase. However, at the higher temperature during the helium shell
flashes the thermal coupling becomes effective.

Because of the sensitivity of the thermal coupling to the temperature, careful
\spro\ calculations have been performed in \cite{Heil08} taking into account
the neutron and temperature profiles in detail. The convective region was
devided into 30\,meshes, and the production and decay of \lu\ was calculated
in each mesh. After each time step of less than one hour, the abundances from
all zones were averaged to take into account convective mixing. Such time
steps are sufficiently short compared to typical reaction rates of the thermal
coupling between \hiK\ and \loK\ states in \lu\ (see Fig.~\ref{fig:rates}).

It was found in \cite{Heil08} that the production ratio between \lu\ and
\hf\ changes dramatically during the evolution of a helium shell flash (see
Fig.~8 of \cite{Heil08}): At the onset of the flash a large ratio between
\hf\ and \lu\ is found because of the dominating production of the \lu\ isomer
in the $^{175}$Lu\rng \lu\ reaction which decays to \hf . As the temperature
during the flash increases, \hf\ is destroyed by neutron capture, but only
weakly reproduced by the decay of the \lu\ isomer because of the thermal
coupling of the \loK\ isomer to the long-living \hiK\ ground state and the
bypass of \hf\ in the subsequent \lu \rng $^{177}$Lu neutron capture
reaction. At the end of the flash neutron density and temperature drop down,
and the initial ratio of \hf\ and \lu\ is almost restored. This means that the
final abundances of \hf\ and \lu\ depend not only sensitively on temperature,
but also on the thermal conditions at the end of the helium shell
flashes. This makes predictions of the \hf\ and \lu\ abundances extremely
difficult and invalidates the simple interpretation of \lu\ as
\spro\ thermometer.

An ideal stellar \spro\ model should be able to reproduce the so-called
overproduction factors
$[N_s(176)/N_s(^{150}{\rm{Sm}})]/[N_\odot(176)/N_\odot(^{150}{\rm{Sm}})]$
(normalized to the $s$-only nucleus $^{150}$Sm) of \lu\ and
\hf\ simultaneously. The slow decay of the long-lived \lu\ ground state in the
interstellar medium prior to the formation of the solar system slightly
reduces the abundance of \lu\ and leaves the more abundant \hf\ almost
unchanged. Thus, overproduction factors of about $1.05 \pm 10\,\%$ for
\lu\ and $1.00 \pm 5\,\%$ for \hf\ are the acceptable ranges
\cite{Heil08}. Such a solution, the so-called ``best case'' with
overproduction factors of 1.04 for \lu\ and 0.95 for \hf , has been found in
\cite{Heil08} with the parameters $b^\ast = 0.022$ and $\tau = 80$\,ps for the
IS at 839\,keV which correspond to an integrated cross section $I_\sigma^\ast
= 71$\,meV\,fm$^2$. Already a slightly increased integrated cross section
(e.g.\ $b^\ast = 0.022$, $\tau = 50$\,ps, $I_\sigma^\ast = 114$\,meV\,fm$^2$)
shifts the overproduction factors to their limits (1.08 for \lu\ and 0.90 for
\hf ) (see Table 7 of Ref.~\cite{Heil08}).

The result of the present study shows that the integrated cross section
of the IS at 839\,keV is roughly one order of magnitude larger than the ``best
case'' of \cite{Heil08} (see Sect.~\ref{sec:stellar} and
Fig.~\ref{fig:rates}). Therefore, the calculations in \cite{Heil08} were
repeated with the larger integrated cross section derived in this work,
i.e.\ with a significantly stronger coupling between the \loK\ isomer and the
\hiK\ ground state in \lu . 

Most of \lu\ is produced in the \loK\ isomer under \spro\ conditions which
decays to \hf . The stronger coupling transforms \lu\ from the \loK\ isomer to
the \hiK\ ground state, thus increasing the overproduction factor of \lu\ and
reducing the overproduction factor of \hf . With the stellar model used in
Ref.~\cite{Heil08} it was not possible to find a 
consistent solution within the given experimental errors of the neutron
capture cross sections of the lutetium and hafnium isotopes and the
uncertainty of the thermal coupling. E.g., using the measured isomeric
production ratio of 0.86 and an integrated cross section $I_\sigma^\ast =
850$\,meV\,fm$^2$ from Sect.~\ref{sec:stellar}, the overproduction factors are
1.80 for \lu\ and 0.61 for \hf\ which is far out of the given range of $1.05
\pm 10\,\%$ for \lu\ and $1.00 \pm 5\,\%$ for \hf\ (see above). 

A further test
with a variation of the neutron production rate of the $^{22}$Ne\ran $^{25}$Mg
reaction within a factor of two was also not successful. (For a detailed study
of the influence of the neutron production rate in the $^{22}$Ne\ran $^{25}$Mg
reaction on the \spro\ nucleosynthesis see \cite{Arl99}.)

The astrophysical ingredients of the stellar \spro\ model have to be known
with very high precision.  Because of the extreme temperature dependence of
the stellar transition rate $\lambda^\ast$ between \hiK\ and \loK\ states in
\lu , a possible solution of the problem may originate from modifications of
the temperature profile or convective mixing during the helium shell
flashes. It has to be noted that an increase of the integrated cross section
$I_\sigma^\ast$ by one order of magnitude (as determined in this work compared
to the ``best case'' of \cite{Heil08}) may be compensated by a minor decrease
in temperature by only about 1.5\,keV leading to the same transition rate
$\lambda^\ast$ via the IS at 839\,keV. Such minor modifications of the
temperature profile will have only small influence on other \spro\ branchings
because most branchings are mainly sensitive to the neutron density. Although
the temperature, which is ``read'' from the \spro\ thermometer \lu , is only
about 1.5\,keV lower than in the temperature profile of the latest
\spro\ study of the $A = 176$ branching \cite{Heil08}, this interpretation of
\lu\ as \spro\ thermometer would by far be too simplistic because of the
extremely sensitive interplay of nuclear and stellar physics in the final
phase of helium shell flashes in AGB stars.

\section{Conclusions}
\label{sec:conc}
It has been shown that the IS at 839\,keV in \lu\ leads to a much stronger
coupling between \loK\ and \hiK\ states in \lu\ than assumed in a recent study
\cite{Heil08}. This result is firmly based on a variety of experimental data
for the IS at 839\,keV. A further enhancement of the coupling may come from
\Kmix\ of two almost degenerate states at 725\,keV; the analysis of the
\Kmix\ has to rely on theoretical arguments up to now \cite{Gin08}. Further
candidates for IS at higher and lower energies do not contribute significantly
to the transition rate $\lambda^\ast$ from \hiK\ to \loK\ states in \lu\ at
\spro\ temperatures. 

From the above results it is obvious that the \spro\ branching at $A = 176$
cannot be well described using the latest \spro\ model \cite{Heil08}. The
nuclear physics ingredients of the model seem to be reliable and based on
experimental data (except the \Kmix\ of the two $7^-$ states at 725\,keV). The
neutron capture cross sections in this mass region have been measured
carefully in the last years including the isomer branch in the $^{175}$Lu\rng
\lu\ reaction \cite{Heil08,Wiss06}. The coupling of \hiK\ and \loK\ states via
IS is known from the combined analysis of all available experimental
data. However, there is still an unsatisfactory large range of allowed values
for the integrated cross section $I_\sigma^\ast$ under stellar conditions
which should be reduced by further experiments (e.g.\ photoactivation using
quasi-monochromatic $\gamma$-rays or using a quasi-stellar photon spectrum
\cite{MohrEPJA}; unfortunately, the relevant energy range is not easily
accessible at the HI$\gamma$S facility \cite{Ton08}). Such experiments may
also address the influence of the suggested \Kmix\ \cite{Gin08} on the
transition rate.

After the nuclear physics input has been considerably improved for the IS at
839\,keV, it was found that the astrophysical interpretation of the \lu
/\hf\ pair as an $s$-process thermometer is again in question due to a strong
overproduction of \lu . The ultimate solution of this problem requires
improved data for the \Kmix\ of the two almost degenerate states at 725\,keV,
and it will further depend on refinements of the stellar physics in the final
phase of helium shell flashes in AGB stars, thus providing deeper insight into
the interesting physics of helium shell flashes.

\begin{acknowledgments}
Encouraging discussions with A.\ Champagne and L.\ Lakosi are gratefully
acknowledged. 
\end{acknowledgments}

\end{document}